\documentclass[aps,prl,reprint,superscriptaddress,longbibliography]{revtex4-1} 
\usepackage[final]{changes} 
\usepackage[utf8]{inputenc}
\usepackage{graphicx}  
\usepackage{dcolumn}   
\usepackage{bm}        
\usepackage{amssymb}   
\usepackage{amsmath}   

\begin{document}


\title{Coexistence of critical sensitivity and subcritical specificity can yield optimal population coding}

\author{Leonardo L. Gollo}%
\affiliation{Systems Neuroscience Group, QIMR Berghofer Medical Research Institute, Brisbane, Australia}
\affiliation{The University of Queensland, Brisbane, Australia}

\begin{abstract}

The vicinity of phase transitions selectively amplifies weak stimuli, yielding optimal sensitivity to distinguish external input.
Along with this enhanced sensitivity, enhanced levels of fluctuations at criticality reduce the specificity of the response.
Given that the specificity of the response is largely compromised when the sensitivity is maximal, the overall benefit of criticality for signal processing remains questionable.
Here it is shown that this impasse can be solved by heterogeneous systems incorporating \emph{functional diversity}, in which critical and subcritical components coexist. 
The subnetwork of critical elements has optimal sensitivity, and the subnetwork of subcritical elements has enhanced specificity. 
Combining segregated features extracted from the different subgroups, the resulting collective response can maximise the tradeoff between sensitivity and specificity measured by the dynamic-range-to-noise-ratio. 
Although numerous benefits can be observed when the entire system is critical, 
our results highlight that optimal performance is obtained when only a small subset of the system is at criticality. 

\end{abstract}

\maketitle


\subsection{1. Introduction}

Reliable estimations of stochastic inputs is a major challenge to many physical~\cite{von56} and biological systems~\cite{rieke99,faisal08}. 
The estimation of input intensity, for instance, is fundamental for life because it constitutes the basis of sensory detection in unicellular organisms~\cite{bray98}, single neurons~\cite{Gollo09}, and animals~\cite{stevens75}. 
A main issue of building a perceptual representation of inputs that arrive in a stochastic fashion derives from their unpredictable nature. 
Hence, the steady state of a discrete random (point) process can be easily confounded with a time-dependent one. 
This is also known as the gambler’s fallacy~\cite{laplace12}, in which the agent (gambler) perceives a steady process as being time-varying 
(with hot and cold moments of luck referring to periods of short and long inter-event intervals~\cite{ayton04}). 
In this case, the perception of a steady mean rate is replaced by erratic fluctuations of the signal around the mean rate. 
In particular,  the larger the amplitude of the fluctuations, the greater the challenge is to overcome this limitation to efficiently estimate the steady rate.

Critical states mediate phase transitions and exhibit numerous special features that distinguish them from other non-critical states: 
They have peaked correlation length, specific heat~\cite{stanley87}, entropy~\cite{haldeman05,mosqueiro13}, information flow~\cite{lizier10,shew11,williams14,vazquez17}, computation~\cite{langton90} and so on. 
An important practical benefit is the maximal dynamic range~\cite{kinouchi06}, which means that critical systems have optimal abilities to distinguish stimulus intensity that spans several orders of magnitude. 
Hence, systems posed at criticality offer maximal sensitivity to detect changes in stimulus rate~\cite{kinouchi06,gollo13,plenz14}. 
This is particularly important to psychophysics in which variations of input must be detected for a wide range of stimulus intensity~\cite{stevens75,kinouchi06}. 
On the other hand, critical systems also exhibit maximal fluctuations~\cite{scheffer09,gal13}. 
This feature strongly compromises the system’s specificity (the confidence of the estimated stimulus intensity). 
Therefore, it is not clear whether the overall advantage of the enhanced sensitivity can overcome the limitations of reduced specificity.

To take advantage of optimal sensitivity, evidences suggest that visual~\cite{shew15,publio09} and auditory systems~\cite{eguiluz00,camalet00,hudspeth08} operate at (or very near) critical states. 
However, it remains puzzling what are the workaround strategies used by living systems to overcome the blatant problem of reduced specificity that is associated with critical states.  
Here we investigate sensitivity and specificity at the critical state in homogeneous and heterogeneous systems. 
We show that heterogeneity in node excitability leads to functional diversity, in which a subgroup of units at criticality coexists with other subgroups of subcritical units. 
This separation can simultaneously confer to networks the benefits of maximal sensitivity of critical units and enhanced specificity of subcritical units. 

\subsection{2. Methods}

For concreteness we employ a general model of excitable networks~\cite{kinouchi06,gollo12, wang17}, adapted to incorporate nodal heterogeneity~\cite{gollo12b}. 
Susceptible (quiescent) nodes can be excited either by (global) external driving or by (local) neighbour contributions: 
External inputs arrive at a steady Poisson rate $h$. 
An input rate $h$ indicates that at each time step $\delta t$ (=1 ms), an external input may arrive with a probability $p_h=1-\exp(-h \delta t)$.   
Neighbour contributions may propagate from active units with a probability $p$.  
Active nodes become refractory in the next time step, and refractory nodes become quiescent with probability $q=0.5$. 
The system is synchronously updated and the time step is fixed at one millisecond. Networks have 5000 nodes that are connected as a sparse Erd\H{o}s-R\'{e}nyi random network with average degree $K=50$. 
Furthermore, diversity is introduced in the excitability threshold $\theta$: Nodes fire from neighbouring contributions if they receive at least $\theta$ inputs within one time step.

\paragraph{Homogeneous networks.}
\added{We consider homogeneous networks in which all nodes have the same excitability threshold $\theta=1$. 
This is a simple yet very influential case that highlights the benefit of criticality to enhance network sensitivity ~\cite{kinouchi06}. 
Other homogeneous networks with $\theta>1$ may also enhance network sensitivity~\cite{gollo12b} but they come along a discontinuous phase transition and hysteresis, increasing the complexity of the dynamics. 
Hence, here we focus on the most standard choice of $\theta=1$ that gives rise to a continuous phase transition~\cite{kinouchi06}. 
}

\paragraph{Heterogeneous networks.}
\added{
Node diversity is a hallmark of the brain~\cite{nelson06} that is associated with several benefits to the system~\cite{mejias12, padmanabhan10, gollo16}. 
For simplicity, diversity is introduced as a discrete uniform distribution of the threshold $\theta=1,2,...,\theta_{max}$. 
The subpopulation of most excitable units has $\theta=1$ as in the previous case of homogeneous populations, and the least excitable subpopulation has $\theta_{max}=6$. 
Integrator units (with $\theta>1$) become active at a lower rate since  
they must integrate several inputs to fire. 
}

\paragraph{Response Function.}
\added{
The mean firing response $F$ across nodes and trials is a smooth sigmoidal function of the input $h$ that vary over orders of magnitude (Fig. 1a). This is called a response (or transfer) function. It indicates the mean output firing rate of the system for varying input rates  (regularly spaced in logarithmic scale). 
Response functions represent a fundamental feature of the system under the rate-code framework~\cite{shadlen94}. 
}

\begin{figure}[!b]
\begin{center}
\includegraphics[angle=0,width=0.96\columnwidth]{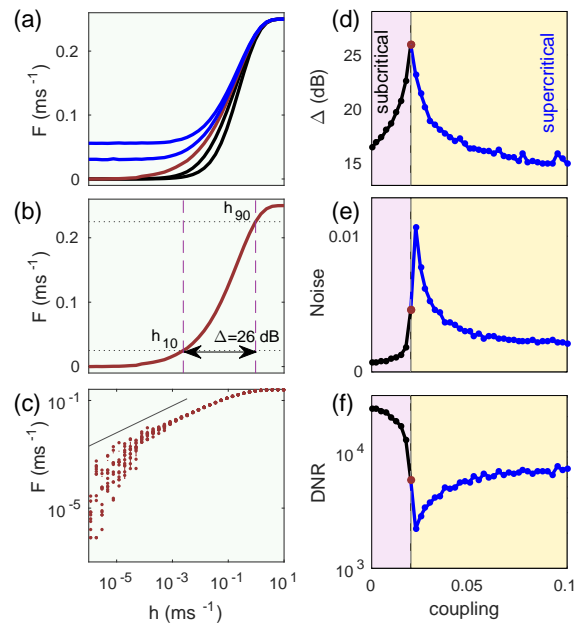}
\caption{\label{homogeneous}  Maximal sensitivity and reduced specificity occurs at criticality in homogeneous networks ($\theta=1$).
  (a)  Family of response functions $F$ for variable input $h$. From right to left: Subcritical curves are in black ($p=0.005$, $0.015$), critical curve is in red ($p_c=0.02$), and supercritical curves are in blue ($p=0.0225$, $0.025$). Response functions represent an average over ten trials.  
  (b)  Illustration of the dynamic range definition for the critical curve. Dashed lines indicate $h_{10}$ and $h_{90}$, whereas dotted lines indicate $F_{10}$ and $F_{90}$.
  (c)  Response function variability for ten independent trials at criticality. Gray line indicates the critical exponent $m=0.5$.  
  (d)  Dynamic range as a function of the coupling strength $p$. 
  (e) Noise, defined as the area between the confidence interval (standard deviation) over the entire response function, versus $p$.
  (f) Dynamic-range-to-noise ratio ($\Delta/noise$) versus $p$. 
  }
\end{center}
\end{figure}

\paragraph{Dynamic range.}
\added{
A key property of response functions (shown in Fig.1a) is the dynamic range~\cite{cleland99}. It measures the range of stimulus intensities resulting in distinguishable network responses. 
The dynamic range quantifies the interval of stimulus in which the network is sensitive to small variations of input, and neglects regions of saturated responses which are too close to the minimal or maximal firing rates (saturation). 
Hence, dynamic range is a measure of network sensitivity that focuses on the range of stimuli in which changes are effectively detected, and influence the firing rate. 
Dynamic range is typically defined as $\Delta=10\, \log_{10}(h_{90}/h_{10})$. In this definition~\cite{kinouchi06}, $h_x\equiv F^{-1}(F_x)$ corresponds to the input level of $F_x=F_0 +\frac{x}{100} (F_{max}-F_0)$, where $F_{max}(=250$ Hz) is the maximal firing rate of the network, and $F_0$ is the minimal firing rate of the network, or the firing rate in the absence of input $F(h=0)$. 
The main elements required to compute the dynamic range of a response function are illustrated in Fig. 1b. 
 }

\paragraph{Noise.}
\added{
Another central feature of response functions is the inter-trial variability, which reflects the specificity of the network response (Fig. 1c). 
 We quantify this \emph{noise} by measuring the area between plus and minus one standard deviation of the mean of the response-function curves. 
 The numerical integration was computed over the entire range of the response function using the trapezoidal rule.  
}

\paragraph{Dynamic-range-to-noise-ratio.}
\added{
As an informative measure of the balance between sensitivity and specificity of the network response, we focus on a simple ratio: 
the dynamic-range-to-noise-ratio (DNR). DNR is applicable and convenient when dynamic range and noise are greater than zero.  Despite the similarity of the DNR with the traditional  \emph{signal-to-noise} ratio (SNR), there are some important differences that need to be emphasised. The noise in our definition of DNR refers to the inter-trial variability of the response (and not to the error associated with the measure of the dynamic range). In addition, in contrast to SNR, the DNR is designed to quantify the trade-off between sensitivity and specificity in response to a range of stimulus intensities covered by the response function. 
Moreover, since uncoupled units (e.g., single neurons~\cite{Gollo09}) may have non-negligible dynamic ranges, the DNR for uncoupled networks may attain large values.
}

\paragraph{Branching ratio.}
\added{
A typical measure of the spreading process in a network is the branching ratio $\sigma$.  
It is defined as $\sigma \equiv \rho(t+1)/\rho(t)$, where $\rho$ is the fraction of active nodes, 
and is computed as the geometric mean over many initial conditions for each value of $\rho$. 
A branching ratio $\sigma>1$ indicates increasing levels of activity; $\sigma<1$ indicates decreasing levels of activity, and $\sigma=1$ indicates stable levels of activity~\cite{kinouchi06}. 
}

\paragraph{Subnetwork perspective.}
\added{
In a heterogeneous network, the branching ratio, the firing rate, and its associated features (dynamic range, noise, and DNR) can be computed for each subnetwork defined by nodes that have the same threshold. 
It is often convenient to analyse the network in this detailed fashion because it clearly highlights the different dynamic behaviour (functional diversity) of the subpopulations. }

\subsection{3. Results}

\paragraph{\added{Sensitivity and specificity in homogeneous networks.}}
In a homogeneous network, nodes require only one neighbouring input to activate. 
This homogeneous case has been subjected to numerous studies since the seminal work of Kinouchi and Copelli~\cite{kinouchi06}. 
Response functions vary depending on the coupling strength (Fig. 1a), and strong coupling allows for self-sustained activity ($F_0>0$). 
Because response functions saturate at a similar level ($h_{90}$, Fig. 1a), the dynamic range  (illustrated in Fig. 1b) is mostly driven by the sensitivity of the system to weak stimuli.  
By varying the coupling strength between nodes $p$, the maximum dynamic range is found at criticality ($p_c=1/K=0.02$). 
This peak shown in Fig. 1d demonstrates the optimal sensitivity of the critical state (red dot)~\cite{kinouchi06}.

Besides from high sensitivity, the critical state also exhibits a large trial-to-trial variability associated with enhanced critical fluctuations. 
\added{The noise curve (computed over the entire response function, see Methods) follows a trend similar to the dynamic range, with a peak close to the critical state and a fast decay away from it (Fig. 1e).} 
The DNR exhibits larger values in the subcritical region, decreases near the critical state, and slowly increases again in the supercritical regime as the coupling is moved away from the trough (Fig 1f). 
The larger DNR at subcriticality indicates an advantage of this regime with respect to the critical and supercritical ones. 
This finding is compatible and also adds to other recent findings and proposals that take advantage of the subcritical regime~\cite{priesemann13, priesemann14, tomen15}. 
Moreover, the minimal DNR occurs very close to the critical state, 
raising concerns about the ability of critical systems to overcome their limited specificity when the sensitivity is optimal. 
Is this an intrinsic and unavoidable limitation, or can it be harnessed by natural or designed systems? 

\begin{figure}[!h]
\begin{center}
\includegraphics[angle=0,width=0.88\columnwidth]{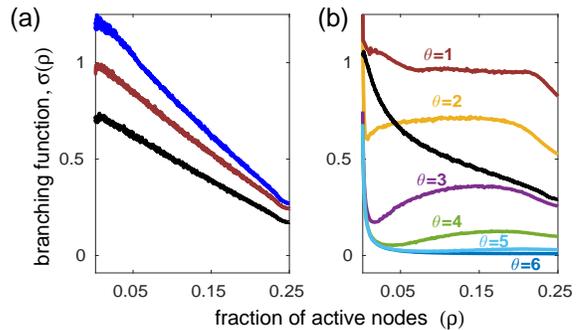}
\caption{\label{branching}  Branching function $\sigma(\rho)$ in homogeneous and heterogeneous systems.
  (a) Branching function decays with the fraction of active nodes ($\rho$). Curves represent $p=0.025$ (blue, supercritical case), $p_c=0.02$ (red, critical), $p=0.015$ (black, subcritical). 
  (b) Branching function in heterogeneous networks with discrete uniform distribution ($\theta=1,2,...,6$) at $p_c^1=0.14$. The branching function of the whole network is in black, and in coloured curves for subnetworks. Results obtained for different initial conditions in the absence of external driving ($h=0$). 
  }
\end{center}
\end{figure}

\paragraph{\added{Branching ratio for different dynamical regimes.}}
To address this question, we first want to highlight the separation of the three non-overlapping regimes occurring in homogeneous systems for different coupling strengths: subcritical ($p<p_c$), critical ($p \approx p_c$), and supercritical ($p>p_c$). 
Each regime has a specific signature of the spreading of activity in the system~\cite{poil08}. 
Since the number of quiescent nodes decays owing to larger numbers of active and refractory nodes, 
the branching ratio also decays with $\rho$ (Fig.~\ref{branching}a).  

In the subcritical state, $\sigma(\rho)$ is consistently less than one, thereby the network activity quickly vanishes in the absence of external input (Fig.~\ref{branching}a). 
At criticality, $\sigma \approx 1$ for small $\rho$, allowing large critical fluctuations that eventually drives the network activity to cease~\cite{Larremore14}. 
Network activity grows ($\sigma>1$) in the supercritical regime for small $\rho$, and stabilises ($\sigma \approx 1$) at a level of self-sustained activity consistent with $F_0$. 
In all cases, nodes behave similarly because they are governed by the same rules. 
Additionally, since the subcritical regime shows a clear enhancement of the DNR (Fig. 1f), 
one possibility to overcome the limitation of the DNR at criticality comes from diversity where some nodes are at criticality and the remainder are subcritical.

\paragraph{\added{Response of heterogeneous networks.}}
Integration ($\theta>1$) makes the transmission of activity less reliable~\cite{wang10}, 
which reduces the branching ratio of these subpopulations (Fig.~\ref{branching}b). 
Compared to the homogeneous case, networks in the presence of diversity require stronger coupling to reach a critical state, which vary for the different subpopulations (multiple-percolation~\cite{colomer14,gollo16}). 
For the critical state ($p_c^1$) of the subpopulation with $\theta=1$, the branching ratio of this most excitable subpopulation is $\sigma \approx 1$ for a large range of $\rho$, and less than one for the other subpopulations. 
While the branching ratio for the whole network (black curve of Fig.~\ref{branching}b) resembles the monotonic decay of homogeneous networks (Fig.~\ref{branching}a), 
the subpopulations exhibit distinct behaviour for the very same coupling strength, which leads to functional diversity. 

\begin{figure}[!h]
\begin{center}
\includegraphics[angle=0,width=0.99\columnwidth]{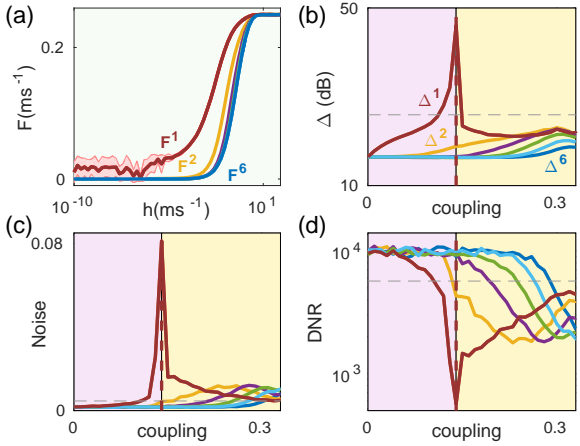}
\caption{\label{het}  Maximal sensitivity and minimal specificity occurs at criticality in heterogeneous networks ($\theta_{max}=6$). 
  (a) Family of response functions $F$ for the different subpopulations $F^{\theta}$ as a function of $h$ for $p_c^1=0.14$. Subpopulations are ordered from left to right. 
  (b) Dynamic range as a function of the coupling strength for the subpopulations $\Delta^{\theta}$. 
  (c) Noise as a function of the coupling strength.
  (d) Dynamic-range-to-noise ratio as a function of the coupling strength. 
  Horizontal dashed lines indicate the corresponding values at criticality for the homogeneous system of Fig. 1.  
  Red dashed lines indicate the critical coupling for the subpopulation of $\theta=1$, $p_c^1$.
  }
\end{center}
\end{figure}

The response functions of the subpopulations in heterogeneous systems are also very different (Fig.~\ref{het}a). 
At $p_c^1$, whereas the subpopulations of integrators ($\theta>1$) are in the subcritical state,
the subpopulation with $\theta=1$ is in the critical state. 
For this critical subpopulation both the dynamic range (Fig.~\ref{het}b) and the noise (Fig. ~\ref{het}c) are enhanced compared to the homogeneous case (shown in Fig. 1). 
In addition, the DNR of this subpopulation essentially replicates the behaviour of the homogeneous case (Fig. 1f): Minimum near criticality, and larger values for the subcritical compared to the supercritical regime (Fig. ~\ref{het}d). 
Notably, in the presence of diversity the dynamic range at criticality for the subpopulation with $\theta=1$ is enhanced. 
Yet, since the noise is enhanced even more for this subpopulation, the DNR is minimal. 
 
\begin{figure}[!h]
\begin{center}
\includegraphics[angle=0,width=0.999\columnwidth]{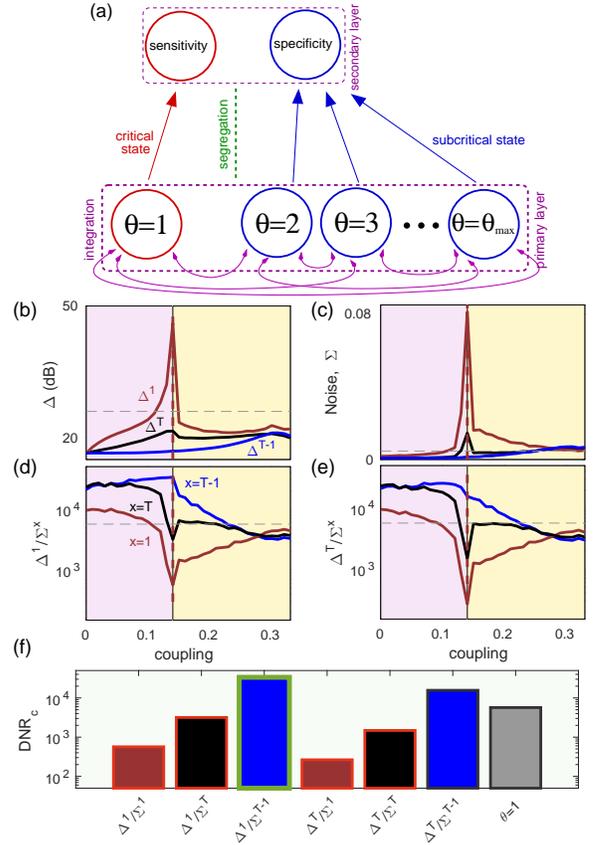}
\caption{\label{comb}  Combining critical sensitivity with subcritical specificity.
(a) Minimal two-level hierarchical structure with horizontal integration (within layer) and vertical segregation (between layers).     
(b) Dynamic range versus coupling strength $p$ for the most excitable subpopulation  $\Delta^{1}$, the average of all subpopulations  $\Delta^{T}$, and the average of all integrators $\Delta^{T-1}$. 
(c) Noise versus $p$ for the same groups: $\Sigma^{1}$ (most excitable, red), $\Sigma^{T}$ (all, black), $\Sigma^{T-1}$ (integrators, blue).
(d) Dynamic-range-to-noise ratio combining the dynamic range from $\Delta^{1}$ with the noise from the other groups.
(e) Dynamic-range-to-noise ratio combining the dynamic range from $\Delta^{T}$ with the noise from the other groups.
     Horizontal dashed lines indicate critical values for the homogeneous system of Fig. 1. 
     Red dashed lines indicate 
     $p_c^1$.
(f) Comparison of the above results for the dynamic-range-to-noise ratio at criticality with the homogeneous case $\theta=1$ (grey). Red borders denote a minimum and the green border denotes a maximum of the DNR at criticality. 
  } 
\end{center}
\end{figure}

\paragraph{\added{Combining critical sensitivity with subcritical specificity.}}
The response of the critical subpopulation has enhanced sensitivity but compromised specificity, and the responses of the subcritical subpopulations have lower sensitivity but improved specificity. 
These specialised responses allow for the possibility of combining the optimal features of the output of each group that coexist in heterogeneous systems (Fig.~\ref{comb}a).  
To explore this avenue,  it is first convenient to split the network into three groups: (i) the subpopulation of most excitable elements ($\theta=1$), (ii) the whole network, and (iii) the union of all integrators ($\theta>1$). These groups exhibit distinct properties with (i) and (iii) showing opposite features and (ii) representing a middle ground between them. 
The most excitable group (i) and the whole network (ii) show peaks for both dynamic range and noise at $p_c^1$. This is  
in contrast to integrators (iii) that exhibit smooth curves (without peaks) for both dynamic range and noise because they are in the subcritical regime (blue lines, Figs.~\ref{comb}b,c). 

By exploring the differences among subpopulations, the DNR is estimated combining the sensitivity of groups (i) and (ii), $\Delta^{1}$ and $\Delta^{T}$, with the specificity of the three groups ($\Sigma^x$, Fig.~\ref{comb}d,e). 
Marrying the dynamic range of group (i) with the noise of group (iii) leads to maximal DNR at criticality (Fig.~\ref{comb}d). 
Thus, optimal sensitivity can coexist with maximal DNR when the estimations of dynamic range and noise come from segregated  groups.

Comparing the DNR at criticality across groups demonstrates the advantage of diversity with respect to homogeneous networks (Fig.~\ref{comb}f). 
This comparison reveals the importance of combining integration and segregation~\cite{tononi94,sporns10} of heterogeneous groups as illustrated in the generic scheme of Fig.~\ref{comb}a: 
Integration occurs within the network as nodes interact with nodes from other subpopulations (purple arrows);  
segregation corresponds to the separation of the outputs from critical and subcritical subpopulations of the primary level towards a secondary level (red and blue arrows). 
Enhanced sensitivity occurs only in one subpopulation ($\theta=1$). 
And optimal DNR requires the dynamic range of the critical subpopulation and the noise of subcritical subpopulations. 
Without this separation, heterogeneous networks perform worse than homogeneous networks 
for dynamic range ($\Delta^T$ in Fig.~\ref{comb}b), noise ($\Sigma^T$ in Fig.~\ref{comb}c), and DNR ($\Delta^T/\Sigma^T$ in Figs.~\ref{comb}e, f). 
To behave optimally the system needs to take advantage of the best feature of each subgroup.

\subsection{4. Discussion}

The importance of critical systems is now widely established~\cite{roli15,mora11, deco12,shew13}, and more recently enhanced consistency and stability of subcritical systems have also been recognised as important features~\cite{priesemann13, priesemann14, tomen15}. 
In addition to affect the dynamics of systems~\cite{cartwright00,tessone06,padmanabhan10} and improve performance in several aspects~\cite{mejias12, gollo16}, 
heterogeneity can also lead to functional diversity in which the dynamics of subgroups are tuned to different dynamical regimes. 
Compared to homogeneous systems, functional diversity can furnish several simultaneous advantages:
(i) Specificity is enhanced in subcritical subpopulations because their response is more reliable and less variable.
(ii) Sensitivity is enhanced in the critical subpopulation because these nodes are more excitable (stronger critical coupling) 
and thus, more effective in amplifying weak stimuli (without significant early saturation for strong stimuli). 
(iii) The ratio between critical sensitivity (dynamic range) and subcritical variability (noise) is maximised. 
Adding to the ever-growing list of advantages of diversity~\cite{van07,joshi09}, we find that functional diversity can promote the coexistence of enhanced specificity with optimal sensitivity and maximal DNR. 

To take advantage of features of critical and subcritical regimes, we propose a simple hierarchical organisation~\cite{zeki88,mesulam98,felleman91,hochstein02, kiebel08} that segregates and integrates the activity of the subpopulations, which are two major processes of complex dynamics~\cite{tononi94,fox12}. 
The proposed structure, illustrated in Fig.~\ref{comb}a, involves integration among heterogeneous elements, 
and segregation of the output from critical and subcritical groups towards higher hierarchical levels. 
At the bottom of the hierarchy, integration only requires recurrent connections, and 
segregation requires the separation of the different responses. 
A future task consists in elucidating the precise mechanism in which neurons at the top of the hierarchy optimally associate inputs from critical and subcritical sources.   
Growing evidence indicates that this separability leads to an exquisite level of specialisation observed in neuronal activity~\cite{quiroga05,gross02}, including regions responding to stimulus noise or prediction error~\cite{iglesias13}. 
Given the specialisation of brain activity, which is most widely found in neuroimaging experiments~\cite{yarkoni11}, 
it is reasonable to expect that the sensitivity and specificity of subnetwork responses are also separable features that can be combined to generate an accurate DNR estimation as well as other more sophisticated combinations of features from heterogeneous responses.

The appealing critical-brain hypothesis has been proposed as a method that would allow the brain to take advantage of several features that are optimised at criticality~\cite{chialvo04,chialvo10,beggs08}. However, despite clear benefits for information processing, this hypothesis remains controversial, especially because numerous evidences of non-critical dynamics also exist~\cite{priesemann13, priesemann14, tomen15}. By revealing that the advantages of critical systems only require a small proportion of critical units, our findings conciliate these two points of view. The coexistence of subcritical and critical subparts of a system can optimise the collective response. Hence, experiments may highlight critical and/or  the subcritical aspects of the dynamics since these regimes are not exclusive.

In summary, we propose a mechanism that allows a critical system to overcome the limitations of enhanced critical fluctuations, improving specificity, optimising sensitivity, and maximising DNR, which reflects a tradeoff between sensitivity and specificity. 
The proposal requires functional diversity and separated pathways that are responsible for conveying complementary representations of the network response: sensitivity from critical elements, and specificity from subcritical ones. 
The benefits of functional diversity are likely (i) to find further applications because they allow for the combination of multiple specialised features, 
and (ii) to transfer to a variety of systems because heterogeneity is ubiquitous in nature.


\paragraph{Funding.}
This work was supported by the Australian Research Council and the Australian National Health and Medical Research Council (APP1110975).


\bibliography{Gollo2017bib_arxiv.bbl}

\end{document}